# Activity Recognition Based on Micro-Doppler Signature with In-Home Wi-Fi

Qingchao Chen[1], Bo Tan [2], Kevin Chetty[1] and Karl Woodbridge[3]
[1] Dept. of Security and Crime Science, [2] Dept. of Computer Science, [3] Dept. of Electronic and Electrical Engineering
University College London, London, UK
{qingchao.chen.13, tan.bo, k.chetty, k.woodbridge}@ucl.ac.uk

**Abstract--** Device free activity recognition and monitoring has become a promising research area with increasing public interest in pattern of life monitoring and chronic health conditions. This paper proposes a novel framework for in-home Wi-Fi signal-based activity recognition in e-healthcare applications using passive micro-Doppler (m-D) signature classification. The framework includes signal modeling, Doppler extraction and m-D classification. A data collection campaign was designed to verify the framework where six m-D signatures corresponding to typical daily activities are sucessfully detected and classified using our software defined radio (SDR) demo system. Analysis of the data focussed on potential discriminative characteristics, such as maximum Doppler frequency and time duration of activity. Finally, a sparsity induced classifier is applied for adaptting the method in healthcare application scenarios and the results are compared with those from the well-known Support Vector Machine (SVM) method.

## I. INTRODUCTION

Life expectancy has increased dramatically over recent decades. Research shows that over 25% of the population is expected to be over 65 years of age by 2051 in the UK. With failing health in later years the monitoring of elderly or disabled people in home environments is therefore becoming a major area of public concern. Indicators such as lack of movement for a period of time, changes in the nature of movement or indication of a fall could be life critical in many cases. The main challenges are:

- Deploying and testing a sensor system which can unobtrusively monitor activities in a residential setting
- Identifying human activities reliably and developing classifiers to identify key movements of interest (e.g. falls, wheelchair upsets)
- Collecting longer term tracking, positioning and activity data
- Integrating such information with other sensor types

Driven by demands from aging and healthcare problem, Ambient Assistant Living (AAL) has been a widely accepted concept whereby various e-healthcare technologies are employed to monitor elderly and disabled people. Within the AAL framework, activity recognition has been an important research topic to facilitate enhanced situational awareness. Activity monitoring sensor technologies including wearables, mobile phones, radio frequency identification (RFID), Passive Infrared (PIR) sensors, ultra-wide bandwidth (UWB) based and vision based sensors are being investigated to detect, recognize and monitor human activity [1]. Among these, sensors embedded in wearable's and mobile phones such as accelerometers and gyroscopes are able to provide some physical information about the subjects, but suffer from low movement update rates of typically less than 5Hz. In addition, people may forget to wear or drop their on-body sensors due to the physical discomfort. PIR sensors are able to only provide the coarse-grained room level existence [2] while RFID based devices employ complex transmitters and receivers, and require pre-planning in order to optimally site the positions of the nodes [1]. Similar to on-body sensors RFID tags or transmitters can also be easily damaged, lost or forgotten [3]. In a similar manner to RFID, UWB activity recognition systems need heavy pre-deployment set up and UWB components are more expensive than other technologies. Video system such as MS Kinect and Intel RealSense have been investigated in some healthcare projects [4]. However, in general, the video camera systems require optimal lighting conditions and the acceptability of deploying video cameras in home environments raises many privacy issues.

In this paper we propose a novel micro-Doppler (m-D) based activity recognition technology using in-home Wi-Fi for pervasive contactless monitoring. Leveraging our passive wireless detection technology developed in [5], we can extract high resolution Doppler information from Wi-Fi signals reflected by personnel as they go about their everyday activities. As the Doppler shift intensity is determined by the speed and direction of a specific movement, a unique Doppler pattern exists corresponding to each class of movement. This is termed the m-D signature and was first investigated by Chen.et.al [6]. M-D signatures can be utilized to differentiate between different types of target and activity, especially human motions, and for distinguishing various types mechanical motions such as those associated with wind turbines and aircraft propeller blades [7], or between bird and drones [8]. Activity recognition in the e-healthcare field has different requirements from the security field. In the security field, classification could be based on multiple cycles of the same motion recorded over a period, however, for some e-healthcare applications, such as fall detection, activities should be recognized instantly so that an alert can be triggered. A monitoring system would therefore ideally operate through a *one-shot classification* approach, where only one cycle m-D signature test sample is utilised for each motion classification.

There have been a number of studies using indoor Wi-Fi access points (APs) as an illuminator to passively detect human movements. The topic of Wi-Fi based human movement Doppler detection was introduced in [5] and [9] extends the capability to Doppler only tracking. Recently, [10] and [11] successfully applied the m-D in the healthcare field for detecting body gestures and respiration. However, these studies focus on improving the Doppler resolution and extending its through-the-wall detection capability, no classification schemes were proposed to classify the gestures and activities that could be used for the e-healthcare applications and health condition analysis. This paper proposes a framework for activity recognition based on m-D signatures using passive Wi-Fi sensing. The work includes a detection scheme, data sample alignment method and application of classifiers, in order to achieve the one-shot classification required. A range of experiments were conducted and an m-D signature dataset involving six motions of interest were collected. The key features of these different signatures are analyzed and the classification results reported and compared with the Support Vector Machine (SVM) classifier.

The rest of the paper is organized as follows: Section II introduces the signal model, Doppler extraction method and general capability of Wi-Fi signals for activity recognition. The novel one-shot classification framework for activity recognition is outlined in Section III. Section IV describes the experiments carried out to collect data and verify the proposed m-D signature classification method. The experimental results and classification outcomes are then described and discussed. Finally, the conclusions drawn from this study and a proposal for further research are presented in Section V.

## II. SIGNAL MODEL AND MICRO-DOPPLER PROCESSING OF PASSIVE WI-FI SENSING

### A. Signal Model and Micro-Doppler Extraction

Passive Wi-Fi radar utilizes the existing Wi-Fi APs as transmitters of opportunity. The signal processing involves cross correlating the reference and surveillance signal channels and using a Fast Fourier Transform (FFT) to find the exact delay $\tau$ and frequency shift $f$ of the strongest signal. This can be represented by the Cross Ambiguity Function (CAF) as follows [12]:

$$CAF(\tau, f) = \int_{-\infty}^{+\infty} e^{-j2\pi f t} ref^H(t-\tau) \times sur(t) dt \quad (1)$$

where $ref(t)$ and $sur(t)$ are Wi-Fi signals from reference and surveillance channels respectively. A complete description of the signal model for the reference and surveillance channel is given in [11]. Additionally, the batch processing of the CAF can be found in [5, 11]. The frequency vector at specific delay induced by the moving target, $X$ is regarded as m-D signature at the specific time. Then signatures are concatenate together as the time-Doppler history signature, which is regarded as the preliminary data to further form the database.

### B. Capability of In-Home Wi-Fi for Activity Recognition

As the velocity of movement dictates m-D signature, it is necessary to analyze the velocity profile of normal motions within-home environments. In general, the velocities associated with sitting down on a chair, falling down and even walking, exhibit a maximum velocity of around 2 m/s [13]. Given that Wi-Fi operates in the 2.4 GHz spectral band this maximum speed limit will induce the maximum Doppler shift of approximately 32Hz. Faced with this small Doppler frequency detection range, a passive Wi-Fi radar could provide very good Doppler resolution for differentiating between various motions based on more signal samples for integration. To summarise, the passive Wi-Fi system is capable of accurately detecting different motions via the Doppler frequency estimation. Although at some specific time instant, the Doppler frequencies from two motions might be the same, there will be a variation in the full temporal Doppler trace describing a particular motion. In general, one activity or one motion will induce a particular velocity-time pattern, which exhibits fruitful features for activity differentiation. In this paper, the concatenated m-D-time history is regarded as the main detected sensor vector for recognition.

## III. ONE-SHOT CLASSIFICATION FRAMEWORK FOR ACTIVITY RECOGNITION

Previously, we have described the signal model for m-D detection using in-home Wi-Fi and discussed the capability of the in-home passive Wi-Fi system for activity recognition based on m-D signatures. In this section, a one-shot classification framework is proposed for activity recognition based on the m-D signature, and includes the following three steps: (i) alignment of m-D data sample structure, (ii) feature selection and (iii) approaches for motion classification.

### A. Align and Adjust Structure of Micro-Doppler Data

To set up a database, the most important requirement is to keep the data samples the same size. There are two dimensions of the m-D signature $X$, which are the frequency bins and the time bins. It is straightforward to maintain the same number of frequency bins through CAF operation by dividing a certain length of the time-domain signal into a fixed number of batches. However, time periods of different motions might be different and the time periods of the same motion are also prone to be different. Therefore, it is difficult to keep the same number of time bins among m-D data samples as it relates to the time periods of motions. In this section, the following two procedures are introduced to set up a well-conditioned database where data samples should first contain the right pattern of the motion, and secondly keep the same size:
- Automatic start and end point detection
- Adjusting the data sample size

To illustrate the two steps clearly, an example of the m-D signature is shown in Fig.1. In Fig.1 (a), the start and end points of an m-D signature are identified by the red arrows. Next in Fig.1 (b), the useful m-D atoms related to the motion are extracted. Finally, to keep the same size of the m-D signature, the data sample is transformed as shown in Fig.1

(b). In the following, the detailed methods to achieve these two goals are introduced.

Here, we propose a method for m-D signature alignment using the standard deviation of frequency bins vector at fixed time bin. Suppose that the m-D frequency vector $X$ is obtained, an intuitive way to justify whether an m-D atom is the start point is to check whether the non-zero frequency bins have large powers. In this way, Constant False Alarm Rate (CFAR) detection might be suitable, but in the indoor environment, as shown in the following Fig.1(b), CFAR detection of m-D signature in the passive Wi-Fi radar might not work well due to the following two reasons:

- The ambiguity peaks in the m-D signature might mislead CFAR to provide wrong decisions, as shown in Fig.1(b).
- Direct Signal Interference (DSI) will generate strong peaks on the zero Doppler line, which might mislead the CFAR detector, as shown in Fig.1(b). It is worth noting that the reason we do not perform the DSI elimination is that the DSI will be an important feature to distinguish different signatures. Details about this point will be described in Section IV.C.

Without elimination of ambiguity peaks and the DSI, another way is to find out some statistical variable that can represent the distribution variations between m-D atoms within and without the m-D signature. The weighted stand deviation might be a good indicator to detect whether an atom contains the m-D signature, as represented by the following two equations:

$$Mean(x) = \frac{\sum_{i=1}^{N} I[i] \times X[i]}{N}, \qquad (2)$$

$$Std(x) = \frac{\sum_{i=1}^{N}(I[i] \times abs(X[i]-Mean(X)))^2}{\sum_{i} I[i]}, \qquad (3)$$

where the vector $I$ is the weights of the frequency bins, which is larger on higher frequency bins. Here we choose the indicator function as $I[i] = i^2$, where $Mean(.)$ is the function to calculate the vector average and $abs(.)$ is the function to calculate its absolute value. In this approach, the starting and ending m-D time bin can be selected once the weighted stand deviations of continuous three time bins are all larger than or smaller than the fixed threshold respectively.

As the m-D data samples has different time bins, we utilize image processing methods to interpolate m-D signature, which means the transformed number of time bins is larger than motion time periods. For the interpolation method, the traditional bi-cubic interpolation is adopted to consider the effect from the neighborhood for interpolating pixel values of unknowns [14]. Finally, we transform each concatenated signature $X_{sample}$ to $X_{inter} \in R^{M \times L}$ with the fixed size of $M$ Doppler bins and $L$ interpolated time bins. Finally, we change the data sample $X_{inter}$ to a vector $d$ with the dimension of $R^{P \times 1}$, where $P = M \times L$ according to equation 4,

$$d = vec(X_{inter}). \qquad (4)$$

B. *Feature Selection Using Principle Component Analysis*

Suppose the database termed as $D \in R^{P \times N}$ is collected, where $P$ is the dimension of each sample (defined in equation 4) and $N$ is the total number of signature samples. To reduce the dimension of dataset and eliminate the noise effect, we apply Principle Component Analysis (PCA) to project the raw dataset onto the subspace spanned by the main eigenvectors. Next we randomly divide it into the dimension reduced training and testing datasets represented as $D_{Red,T}$ and $D_{Red,S}$ respectively. Here, it should be noted that we used the eigenvectors of the training set to project the test samples, as it is assumed that we cannot know all the test samples a priori.

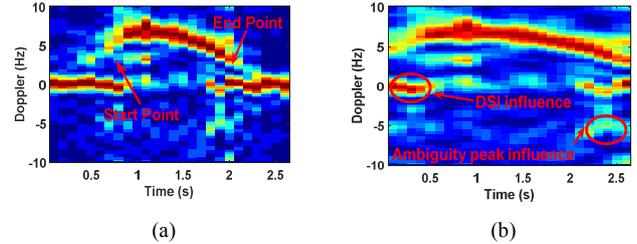

(a) (b)

Fig. 1 (a) Start and End point illustration; (b) The useful part of the m-D signature is adjusted into the same size, also the DSI influence and the ambiguity peak influence.

C. *Sparse Representation Classifier in Micro-Doppler Signature Classification*

The key idea behind the sparse representation classifier (SRC) is the discriminative power of sparse representation which chooses the most compact representation instead of the less compact ones [15]. Intuitively, there are multiple solutions to represent the test sample by linear combination of the training samples and not all the representations can help the classification. However, SRC guarantees that the test sample can be represented by a very small portion of the training samples. Due to this sparsity of linear combination weights, a test sample can be easily classified.

Next, we utilize a sparse signal recovery framework to find labels of test data samples. Suppose we have obtained a test sample $y \in R^{Md \times 1}$ from the set $D_{Red,S}$, if a small number of samples in the training dataset can be utilized to represent the test sample $y$ with minimum residuals, then the training data samples with the bigger supports might belong to the same class as the test sample. Using this principle results in the following optimization problem is proposed:

$$\arg\min_s \|y - D_{Red,T} \times s\|_2^2 + \|s\|_0, \qquad (5)$$

where $s$ is the sparse coding vector and $\|s\|_0$ is the zero norm operator defined as the number of non-zero elements in the vector. Several methods have been utilized to solve this optimization problem, for example, the L1 solver [16] or the Orthogonal Matching Pursuit (OMP) [17]. However, the L1 solver is computationally expensive and the OMP is noted for its slow convergence and inaccuracy [18]. Therefore, we choose the subspace pursuit (SP) as it has a faster convergence rate than the L1 without loss of accuracy [18].

The classification task is operated based on the reconstruction error using the $i^{th}$ class samples. Specifically, reconstruction error of the $i^{th}$ class can be calculated by subtracting the linear combinations of atoms from $i^{th}$ class, as indicated by equation 6. The test sample label can then be classified by looking for the minimum of the reconstruction error among all classes, as the following:

$$\arg\min_i \|y - D_{\text{Re}d,T} \times x_i\|_2 \quad (6)$$

## IV. EXPERIMENTS AND ACTIVITY RECOGNITION RESULTS

To test the classification scheme proposed in Section III, various experiments were designed and data were collected through a campaign in a home-based test site. The passive Wi-Fi system and test sites used are introduced in section A and B respectively. In section C the m-D database collected, containing six different activities is described. In Section D the results are discussed and analyzed.

### A. System Design and Implementation

The Wi-Fi passive radar utilizes the Software Defined Radio (SDR) system. As shown in the following Fig.2, three Universal Software Radio Peripheral (USRP) N210s are synchronized with the Octo-Clock device and three patch antennas (10 dBi gain and 20 degree beam-width) are utilized to collect Wi-Fi signals centered at 2.462GHz. After sampling and transferring signal to the PC (DELL-M4700 laptop with Intel Core i7-3940XM CPU at 3GHz), the CAF is processed and we display and record the real-time Doppler-time history into the text files. For the Wi-Fi AP, we are using the Edimax 300 Range Extender, with two Omni-directional antennas of 3dBi gain.

The processing scheme uses real-time batched processing CAF [5], with a sampling rate of 2MHz and batch number of 20 but zero padding to 50. The overlapping time is 0.04 second and the integration time is set up to 0.4 second. The signature samples are collected real-time into the PC text files and they are aligned to 51 Doppler bins and 50 time bins.

### B. Experimental Tests Design

Besides testing the classification framework proposed in previous section, the experiments were conducted in the house alongside other SPHERE sensors, including an accelerometer and video monitoring. It is planned that these sensor data can be integrated and correlated with our passive Wi-Fi data in the future.

As shown in Fig.3, the experiment is conducted in a living room of a house with 3.87 meter by 3.67 meter. The position of reference antenna is at the height of 1.3 meters pointing towards the AP, while the surveillance antennas are located on the ground and pointing up to testing position with the angle of 45 degree. During the experiments, four male targets were standing in the same testing position and six motions were recorded using the passive Wi-Fi radar.

### C. Overview of Micro-Doppler Signatures and Analysis

In this section, six m-D signatures corresponding to activities listed in Table.1 are shown as the following Fig.4.

As shown in Fig. 4, visually the six m-D patterns exhibit different patterns and the main characteristics to distinguish them can be summarized as follows:
1. The maximum Doppler shift
2. Time duration of the m-D signature
3. Does Doppler frequency ranges from negative to positive or just negative/positive
4. Whether a strong zero Doppler line caused by the DSI or multipath occurs in the m-D signature.

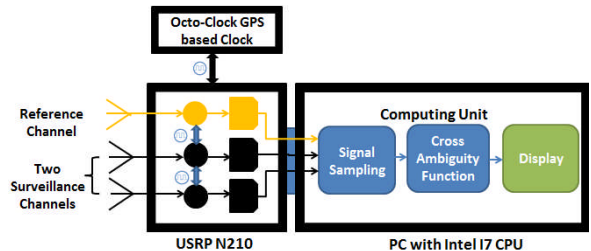

Fig. 2 Passive Wi-Fi System Architecture

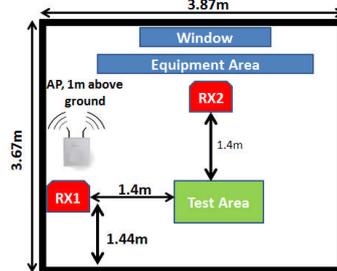

Fig. 3 Passive Wi-Fi Radar Experiment Geometry.

In general, the maximum Doppler frequencies of these six m-D signatures range from 2.5Hz to 4.5Hz and motions with different maximum Doppler frequencies will help distinguish the signatures. The second discriminative feature relates to the relative direction of motion, indicated by the sign (positive or negative) of the Doppler frequency: some motions induce Doppler frequency that goes from negative to positive, (e.g. motions 2 and 3), while others induce only positive or negative Doppler frequencies (motions 4, 5 and 6). Although motion 2 and 3 both have the similar patterns (from negative to positive), the time duration of each signature segmentation increases the discrimination, such as the shorter duration of positive Doppler frequency in [motion 2, channel 1] than the positive Doppler frequency in [motion 1, channel 1]. The final feature that might be distinguishable is whether the zero Doppler line exists during the motion. A clear example is the comparison between the [motion 2, channel 1] and the [motion 6, channel 1], where the Doppler signature patterns are similar, but the former has a strong zero Doppler line while letter does not. The reason why [motion 6, channel 1] exhibits no zero Doppler line is when the target gets out of the bed, the bulk motion blocks the direct signal to the receiver 1.

For m-D classification, these empirical features agree closely with the intuitive visual interpretation. However, obtaining these features requires complex feature selection methods which are prone to be erroneous and have a big influence on the classification outcome. It is clear that it will always be difficult to fully represent a high-dimensional dataset using just four to six empirical features. Recently, the eigenvector based features in both time and Doppler directions by Singular Value Decomposition (SVD) have become popular. However, the SVD requires multiple m-D signatures as a testing data sample for feature extraction. It seems that SVD does not fit our one-shot classification requirement for healthcare applications. In a real scenario, we clearly cannot perform classification after a target has fallen down multiple times. For our classification scheme, we utilize the raw training m-D data (or the reduced-dimension data vector) to represent the test samples, which avoid the erroneous empirical feature selection and the unrealistic requirements of multiple cycles of motion in the test samples. In our assumption, during the representation using the SRC classifier, these features have already been considered.

In addition, the design of our three-receiver synchronized system is to increase the coverage of test scenarios, so that the motions from most of the directions can be detected. In the experimental layout used in these tests, the Channel 1 receiver is close to the AP transmitter which results in a quasi-monostatic geometry rather than the bistatic perspective seen by the channel 2 receiver. This results in a generally smaller power return and Doppler shift in Channel 2 and hence some differences in Doppler characteristics being observed between the two channels.

### D. Classification Results and Analysis

In this section, we show the classification results using the SRC and the SVM classifiers for comparisons. We randomly selected 40% of the samples as training and the others as test samples.

In general, SRC and SVM utilize different frameworks for classification. The SVM first identifies the support training samples with the largest marginal distance to other classes and then directly measures the distance between the supporting vectors and the test samples. The SRC however doesn't measure the distances directly but represents the test samples using linear combinations of training datasets. To handle the coherence of the dictionary, a zero-norm constraint is used (defined in equation 5), so that the training samples used to represent the test data should be as sparse as possible. This sparsity constraint is advantageous as the misleading samples (around boundary of two classes) may not to be chosen as the compact basis. From Table 2, SRC seems to outperform SVM and the intuitive reason is: the support training samples might be misleading due to inter-class similarity but SRC avoids them as it selects the most compact basis to represent the test sample. This compact representation fits into the following requirement in healthcare monitoring field: flexibility to support new users without need to re-training the system [19], because the sparsity level is under control and easy to adjust (no need to perform the long-time re-training) and it will contribute directly to reduce the inter-class similarities.

Through observations on the confusion matrix in Table 3, 8% of motion 2 samples are misclassified to motion 6 because the similar negative Doppler frequency (shown in Fig.4). Another reason might be that the duration of positive Doppler frequency in [motion 2, channel 1] is short and close to the zero Doppler line, so that it has relatively small influence onto the reconstruction error in the classification results. Motion 5 and motion 6 are prone to be misclassified with each other and the reason is obvious: they exhibit similar Doppler patterns. Actually, these two motions are similar as they both are getting out of the bed.

Table.1 List of Motions to Recognize

| Motion Index and NO. of Samples | Description (At the fixed testing position.) |
|---|---|
| M1 (40) | Subject picks up from the ground and stand up. |
| M2 (40) | Subject sits down on a chair. |
| M3 (40) | Subject stands up from a chair. |
| M4 (10) | Subject falls down onto the mattress. |
| M5 (20) | Subject stands up after falling. |
| M6 (20) | Subject lies on a mattress first then gets out of it. |

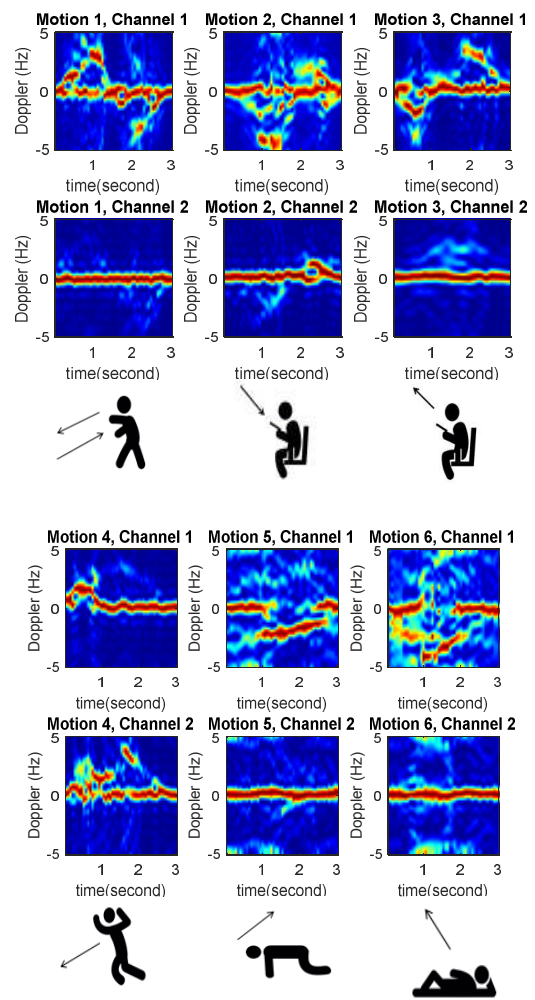

Fig.4 Motion 1 to 6, m-D signature from two channels. Note: the matrix value in the signature is normalized from 0 to 1.

An interesting point is the classification result of motion 4 (falling down). Due to the uncontrollable characteristics of falling down during experiments, it seems that receiver 2 picks up the stronger Doppler frequency. As a result of the stronger power of Doppler signatures, the classification accuracy from channel 2 is much higher than the ones from channel 1. This interesting result leads us to consider the introduction of jointly sparsity based classifiers in the future, which can be easily adapted from the SRC and are based on fusion of multiple sensors information. This method can increase coverage and classification accuracy even if one channel sensor is not working properly [20]. Although some similar Doppler characteristics have been observed in previous studies we believe this is the first report of detailed classification of Healthcare related activities and shows significant promise for application in an automated monitoring and alert system.

Table 2 Classification Results of Channel 1 and Channel 2

| Ch.1 | M1 | M2 | M3 | M4 | M5 | M6 | AVG |
|---|---|---|---|---|---|---|---|
| SRC | 100% | 91.7% | 91.7% | 100% | 83.3% | 67% | 90.2% |
| SVM | 63.6% | 33.3% | 83.3% | 0% | 0% | 20% | 46% |
| Ch.2 | M1 | M2 | M3 | M4 | M5 | M6 | AVG |
| SRC | 91.7% | 91.7% | 91.7% | 33% | 83.3% | 74% | 85.2% |
| SVM | 63.6% | 33.3% | 83.3% | 0% | 0% | 20% | 46% |

Table 3 Classification Confusion Matrix using SRC

| | | Classified Results | | | | | |
|---|---|---|---|---|---|---|---|
| | | M1 | M2 | M3 | M4 | M5 | M6 |
| Ground Truth | M1 | 95.8% | 0% | 0% | 0% | 0% | 0% |
| | M2 | 0% | 91.7% | 8.3% | 0% | 8.3% | 0% |
| | M3 | 0% | 0% | 91.7% | 33% | 0% | 0% |
| | M4 | 4.1% | 0% | 0% | 67% | 0% | 0% |
| | M5 | 0% | 0% | 0% | 0% | 83.33% | 29.2% |
| | M6 | 0% | 8.3% | 0% | 0% | 8.3 % | 70.8% |

## V. CONCLUSION

In this paper, it is demonstrated that important activities related to healthcare monitoring can be recognized unobtrusively with high detection and classification rates, and at low-cost using in-home Wi-Fi. Specifically, a framework for activity recognition has been designed and tested utilising m-D signatures obtained from our passive Wi-Fi radar prototype system. This includes new signature detection, data sample alignment and classification schemes. This framework does not utilize the conventional empirical features but employs sparsity induced whole matrix classification scheme to fit the one-shot classification required by e-healthcare applications.

Various experiments involving six key activities of interest in the e-health field were conducted and the classification results were compared with SVM. Our sparsity based classifiers are demonstrated to outperform the SVM in the healthcare environment context and provide a feasible tool for real time healthcare alerts. This system also has potential flexibility to support new users and capability to increase classification accuracy and coverage by using joint sparsity based classifiers and fused multi-sensor data.

## ACKNOWLEDGEMENTS

This work was performed under the SPHERE IRC and funded by the UK Engineering and Physical Sciences Research Council (EPSRC), Grant EP/K031910/1.